\documentclass[%
 reprint,
 amsmath,
 amssymb,
 pre
]{revtex4-1}
\usepackage{color}
\usepackage{multirow}
\definecolor{blue}{rgb}{0,0,0}
\usepackage{graphicx}
\usepackage{dcolumn}
\usepackage{float}
\usepackage{bm}
\begin{document}

\preprint{APS/123-QED}

\title{Chaos-based Wireless Communication Resisting Multipath Effects}

\author{Jun-Liang Yao}
\author{Chen Li}%
\author{Hai-Peng Ren}%
\affiliation{Shaanxi Key Laboratory of Complex System Control and Intelligent Information Processing, Xi'an University of Technology, Xi'an 710048, China}
\author{Celso Grebogi}
\affiliation{Institute for Complex System and Mathematical Biology, University of Aberdeen AB24 3UE, United Kingdom}
%
%
%

\date{\today}

\begin{abstract}
In additive white Gaussian noise channel, chaos has been shown to be the optimal coherent communication waveform in the sense of using a very simple matched filter to maximize the signal-to-noise ratio. Recently, Lyapunov exponent spectrum of the chaotic signals after being transmitted through a wireless channel has been shown to be unaltered, paving the way for wireless communication using chaos. In wireless communication systems, inter-symbol interference caused by multipath propagation is one of the main obstacles to achieve high bit transmission rate and low bit-error rate (BER). How to resist the multipath effect is a fundamental problem in a chaos-based wireless communication system (CWCS). In this paper, a CWCS is built to transmit chaotic signals generated by a hybrid dynamical system and then to filter the received signals by using the corresponding matched filter to decrease the noise effect and to detect the binary information. We find that the multipath effect can be effectively resisted by regrouping the return map of the received signal and by setting the corresponding threshold based on the available information. We show that the optimal threshold is a function of the channel parameters and of the information symbols. Practically, the channel parameters are time-variant, and the future information symbols are unavailable. In this case, a suboptimal threshold is proposed, and the BER using the suboptimal threshold is derived analytically. Simulation results show that the CWCS achieves a remarkable competitive performance even under inaccurate channel parameters.
\end{abstract}

\pacs{Valid PACS appear here}
\maketitle
\section{Introduction}
In recent years, the rapid development of wireless communication technology \cite{Key-1,Key-2} has dramatically changed the way we live. As compared with wired communication, the wireless communication channel is more complicated \cite{Key-3}. The wireless channel constraints include limited bandwidth, multipath propagation, Doppler shift and complicated noise contamination, etc. Firstly, the inter-symbol interference (ISI) caused by multipath propagation increases bit-error rate (BER). Secondly, the Doppler shift produced by mobile terminals and scattered clusters leads to time-variant channel properties. Wireless communication systems have to deal with these problems effectively.

Chaos has attracted lots of attention in the communication field \cite{Key-4,Key-5,Key-6,Key-7,Key-8,Key-9,Ren2017A,Hai2017Secure} since 1990, due to its intrinsic properties suitable for communication applications \cite{Key-10}, such as broadband, orthogonality, easy to generate, and pseudo-randomness, etc. The maximum rate of communication with chaos was reported in \cite{Key-4,Key-11}. The channel capacity of chaos-based digital communication in noisy environment was given in \cite{Key-12} and the method for filtering noise in chaotic signal was devised in \cite{Key-13}. Chaotic signals were also used as modulation signals for coherent and non-coherent communication \cite{Key-14}. {An integrated chaotic communication scheme based on chaotic modulation was presented in \cite{Baptista:2000}}. Since chaos was reported to be successfully used in an optic fiber communication system in order to get higher bit transmission rate \cite{Key-15}, the research on communication with chaos has been focusing on practical communication channels. The wireless channel is a practical channel with complicated constraints. Among them, the limited bandwidth had been considered in Ref. \cite{Key-16} to show that the BER increased significantly if the bandwidth of the communication channel is low. {The effects of bandwidth on the chaotic synchronization was investigated, and a synchronization method for bandlimited channels was proposed in \cite{Eisencraft:2011}}. Some techniques for improving the performance of chaos-based communication systems in non-ideal channels were reviewed in \cite{Eisencraft:2012}, and some information recovery methods were proposed to decrease the BER in limited bandwidth case \cite{Key-17}. Recent works in \cite{PhysRevLett.97.024101,Key-18,Erratum,Key-19,Key-20} have shown that chaotic signals are optimal waveforms for coherent communication in the presence of noise in the sense of using simple matched filter to maximize the signal-to-noise ratio (SNR). The proposed chaotic waveforms have been successfully used in optical communication \cite{Jiang:16} and digital communication \cite{Bailey:2015npa,Ren2017A}. The wireless communication channel constraints make the chaotic signal transmitted in it change dramatically \cite{Key-21}. A fundamental problem is then whether the information in a chaotic signal is lost after being transmitted through a wireless channel. Reference \cite{Key-22} gave an answer to this problem, and showed that the information is not lost.

\begin{figure*}[t]
\centering
\includegraphics[scale=0.62]{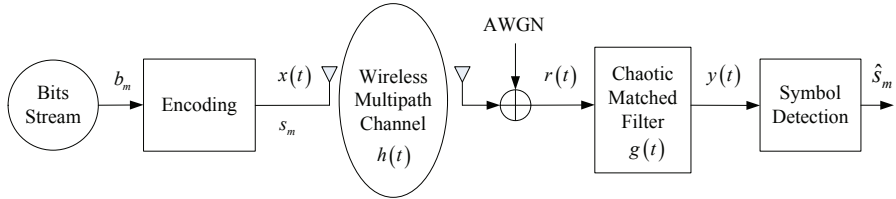}
\caption{Schematic diagram of the chaos-based wireless communication system.}
\end{figure*}

A major issue is to understand and to deal with ISI and time-variant channel parameters, caused by multipath propagation and Doppler shift in the chaos-based wireless communication system (CWCS). In a conventional wireless communication system, the known channel parameters allows the receiver to tackle ISI by using channel equalization techniques. However, the equalization algorithms are complicated and their performance is susceptible to noise.

In this work, we show that the multibranch return map, produced by multipath propagation in the CWCS, can be regrouped according to transmitted symbols. The multipath effects can be hindered by setting a proper detection threshold for the different groups. The optimal threshold is obtained if all symbols and channel parameters are known in advance, and an suboptimal threshold is obtained if the future symbols are unknown. The closed-form expressions for BER of CWCS using the optimal threshold and the suboptimal threshold are derived in this work. The simulation results demonstrate that the proposed CWCS is robust to channel errors. In fact, in the proposed CWCS, a simple matched filter is used to maximize the SNR, which replaces the complicated conventional matched filter algorithm with an even simpler algorithm. More importantly, a suboptimal threshold can be used to relieve the multipath effects in the proposed method, which is simpler and better as compared to the traditional channel equalization methods. Thus, CWCS might be a competitive alternative for a conventional wireless communication system.

This paper is organized as follows. Section II presents the CWCS and the method for resisting multipath effects. Simulation and analysis results are reported in Sec. III to verify the validity of the proposed method. Finally, some concluding remarks are given in Sec. IV.

\section{{Chaos-based wireless communication system}}
A point-to-point CWCS, as shown in Fig. 1, is considered in this work. At the transmitter side, a binary bit sequence $b_m\in[0,1]$ is encoded into the analog signal $x(t)$ by using a hybrid system given by Eq. (1) \cite{Key-18} and by using the encoding method in \cite{Key-23,Key-24,Key-25}:
\begin{equation}\label{Dynamic}
\left\{
\begin{aligned}
&\ddot{x}-2\beta\dot{x}+(\omega^2+\beta^2)(x-s)=0 \\
&\dot{x}(t)=0 \Rightarrow s(t)={\rm{sgn}}(x(t)),\\
\end{aligned}
\right.
\end{equation}
with parameters $\omega=2{\pi}f$ and $0<\beta{\le}f\cdot$ln2, where $f$ is the base frequency. The variable $s={\rm{sgn}}(x)$ switches its value when $\dot{x}=0$, keeping its value at other times. It is worth noting that the system of Eq. (\ref{Dynamic}) with $\beta=f\cdot$ln2 has no grammar restrictions, and the system could be controlled to generate any symbol sequences using small perturbations. The exact analytic solution of Eq. (1) is $x(t)=\sum_{m=-\infty}^{\infty}s_m{\cdot}p(t-m/f)$, where $s_m\in[-1,1]$ is the bipolar symbol representing information bit $b_m$. The basis function is given by
\begin{equation}\label{pt}
p(t)=\left\{
\begin{aligned}
&(1-e^{-\beta/f})e^{{\beta}t}({\rm{cos}}{\omega}t-\frac{\beta}{\omega}{\rm{sin}}{\omega}t),\ (t<0) \\
&1-e^{\beta(t-1/f)}({\rm{cos}}{\omega}t-\frac{\beta}{\omega}{\rm{sin}}{\omega}t),\ \ (0\le{t}< 1/f)\\
&0,\qquad \qquad \qquad \qquad \qquad \qquad \quad \quad (t\ge 1/f).
\end{aligned}
\right.
\end{equation}

{\color{blue}
At time $t$, the encoded signal can be represented as
\begin{equation}\label{xt}
\begin{aligned}
x(t)=s_n+&\Big\{-s_n+(1-e^{-\beta/f})\sum_{i=0}^{\infty}s_{i+n}e^{-i\beta/f}\Big\}\\
&\times e^{\beta(t-n)/f}\Big({\rm{cos}}{\omega}t-\frac{\beta}{\omega}{\rm{sin}}{\omega}t\Big),
\end{aligned}
\end{equation}
where $n$=floor$(ft)$  is the integer part of time $ft$, $s_n$ is the sampling value of $s(t)$ at time $t=n/f$. By sampling the time series given by Eq. (\ref{xt}) with the interval $T_s=1/f$, we have:
\begin{equation}\label{xn}
x_n=e^{n\beta/f}\Big\{x_0-(1-e^{-\beta/f})\sum_{i=0}^{n-1}s_ie^{-i\beta/f}\Big\},
\end{equation}
where $x_n$ is the sampled value of $x(t)$ at the time $t=n/f$. From Eq. (\ref{xn}), we have:
\begin{equation}\label{x0}
x_0=e^{-n\beta/f}x_n+(1-e^{-\beta/f})\sum_{i=0}^{n-1}s_ie^{-i\beta/f}.
\end{equation}

If $n\rightarrow\infty$, it yields the relationship between the initial value $x_0$ and the future symbols (information) $s_i\ (i=0,1,\cdots,\infty)$ given as:
\begin{equation}\label{x02}
x_0=(1-e^{-\beta/f})\sum_{i=0}^{\infty}s_ie^{-i\beta/f}.
\end{equation}

In other words, from Eq. (\ref{x02}), all the future symbols $s_i$ are encoded by an initial value. Practically, limitations due to parameter accuracy or numerical solution error, etc., we cannot encode all future symbols in one initial condition. We use
\begin{equation}\label{xej}
x_e(j)=(1-e^{-\beta/f})\sum_{i=jN_c+1}^{(j+1)N_c}s_ie^{-(i-1)\beta/f},
\end{equation}
where $N_c$ is the time interval to reset the initial condition for system (\ref{Dynamic}). This means that, by resetting the state of system (\ref{Dynamic}) as $x(jN_cT_s)=x_e(j)$ given by Eq. (\ref{xej}) at time $t=jN_cT_s$, we can encode future $N_c$ symbols as given by $s_i\ (i=jN_c+1,...,(j+1)N_c)$. This is the general idea to encode information in our method. There are some details of the technique which can be found in \cite{Key-23,Key-24}.
}

After encoding, the chaotic signal $x(t)$ is transmitted through a wireless channel given by

\begin{equation}\label{ht}
h(t)=\sum_{l=0}^{L-1}\alpha_l\delta(t-\tau_l),
\end{equation}where $\alpha_l$ and $\tau_l$ are the attenuation and propagation time delay corresponding to path $l$ from the transmitter to the receiver, and $\delta(\cdot)$ is the Dirac $\delta$ function. The received signal is
\begin{equation}\label{rt}
\begin{aligned}
r(t)&=h(t)*x(t)+w(t)\\
&=\sum_{l=0}^{L-1}\alpha_lx(t-\tau_l)+w(t)\\
&=\sum_{l=0}^{L-1}\alpha_l\sum_{m=-\infty}^{\infty}s_mp(t-\tau_l-m/f)+w(t),\\
\end{aligned}
\end{equation}where `$\ast$' denotes convolution and $w(t)$ is an additive white Gaussian noise (AWGN). Assume that the delay $\tau_l\ (l=0,1,\dots,L-1)$ satisfies $0=\tau_0<\tau_1<\cdots<\tau_{L-1}$ and that the attenuation $\alpha_l$ is modelled as an exponential decay $\alpha_l=e^{-\gamma\tau_l}$ with damping coefficient $\gamma$ {\cite{Pekka:2009}}. Due to the form of $p(t)$ in Eq. (2) and the time delay $\tau_l$, the received signal $r(t)$ depends not only on the current and future symbols, but also on the past symbols.
\begin{figure*}[tbh]
\centering\includegraphics[scale=0.6]{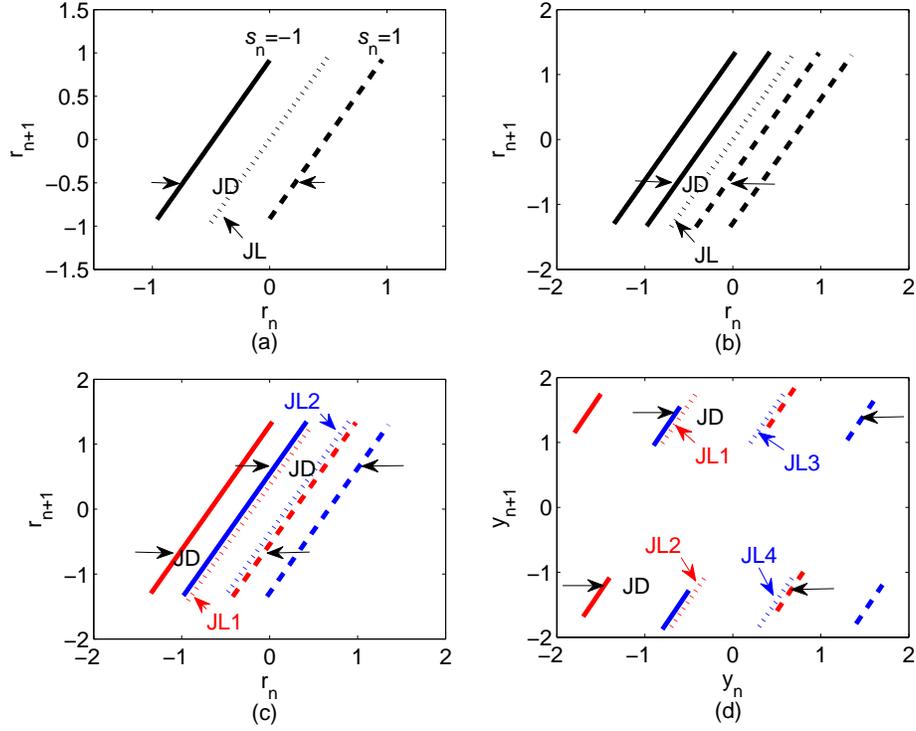}\caption{Return maps for different cases ($f=1,\beta=0.65$): (a) return map of $r(t)$ over the single path channel, (b) return map of $r(t)$ over two-path channel with $\tau_0=0,\tau_1=1,\gamma=0.9$, (c) regroup of return map in Fig .2(b), (d) return map of filtered output signal, $y(t)$, over two-path channel with $\tau_0=0,\tau_1=1,\gamma=0.9$.}
\end{figure*}

By sampling $r(t)$ at frequency $f$, and recording it as $r_n=r(n/f)$, the return map of $r(t)$ can be given by
\begin{equation}\label{rn}
\begin{aligned}
 r_{n+1}=e^{\frac{\beta}{f}}r_n-(e^{\frac{\beta}{f}}-1)\sum_{l=0}^{L-1}\alpha_ls_{n+\lceil{-\tau_lf\rceil}}.
\end{aligned}
\end{equation}
where $\lceil{-\tau_lf\rceil}$ represents the ceiling integer of $-\tau_lf$. Equation (\ref{rn}) shows that the return map of $r(t)$ is related to the current symbol $s_n$ and the past symbols $s_{n+\lceil{-\tau_lf\rceil}}$.

For a single path channel, the return map of $r(t)$ is shown in Fig. 2(a), where there are two branches plotted, shown as a black solid line and black dashed line, respectively. The black solid line corresponds to $s_n=-1$, and the black dashed line corresponds to $s_n=1$. At the $n$th sampled time, the information symbol $s_n$ can be detected by comparing $r_n$ with a judgment threshold $\theta_n$. Then `$s_n=-1$' if `$r_n\le\theta_n$' and `$s_n=1$' if `$r_n>\theta_n$'. In fact, we can find a $judgment\ line$ (JL), as shown in Fig. 2(a), to distinguish the two branches of the return map. We define the horizontal distance between the two branches as the $judgment\ distance$ (JD), which is shown by the distance between the two black arrows in Fig. 2(a). In the single path channel, JD$=2(1-e^{-\beta/f})$. The BER depends on both the JD and JL. In Fig. 2(a), the black dotted line, $r_{n+1}=e^{\beta/f}r_n$, is the optimal JL for distinguishing `$s_n=1$' and `$s_n=-1$', and the corresponding threshold is $\theta_n=e^{-\beta/f}r_{n+1}$.

The return map of $r(t)$ over two-path channel with propagation delay $\tau_0=0$, $\tau_1=1$ is given in Fig. 2(b), where additional branches appear due to multipath propagation. In fact, the branch number is equal to $2^L$, where $L$ is the multipath number. In Fig. 2(b), the black solid lines and the black dashed lines represent `$s_n=-1$' and `$s_n=1$', respectively, as in Fig. 2(a), but in this case JD=$2(1-e^{-\beta/f})(1-e^{-\gamma})$. Increasing bit error detection happens because JD is decreased.

Our finding is that these branches, caused by multipath, can be regrouped into different pairs once the past symbols $s_{n+i} (i=\lceil{-\tau_1f\rceil},\ \lceil{-\tau_2f\rceil}, \cdots,\lceil{-\tau_{L-1}f\rceil})$ are determined. For example, for $L=2$, the branches in Fig. 2(b) can be regrouped as shown in Fig. 2(c), where one past symbol $s_{n-1}$ is needed to regroup the branches. If $s_{n-1}=-1$, the branches pair corresponds to the red pairs, else (i.e., $s_{n-1}=1$) the branches pair corresponds to the blue pairs, while the solid lines represent `$s_n=-1$' and the dashed lines represent `$s_n=1$' as well. Note that the branches in Fig. 2(c) are the same as that in Fig. 2(b), but after regrouping the branches that confuse us, we can use separated JLs, so that JD is the same as that for the single path channel, i.e., $2(1-e^{-\beta/f})$. In order to have the lowest BER, we use the optimal JLs, $r_{n+1}=e^{\beta/f}r_{n}+(e^{\beta/f}-1)e^{-\gamma}s_{n-1}$, plotted using the dotted lines with different colors for the different group in Fig. 2(c), where the red dotted line corresponds to the red group (for the case of $s_{n-1}=-1$) and the blue dotted line corresponds to the blue group, (for the case of $s_{n-1}=1$). The thresholds is $\theta_{n}=e^{-\beta/f}r_{n+1}+(1-e^{-\beta/f})e^{-\gamma}s_{n-1}$. For more paths, we just need  more past symbols to regroup and derive the optimal JLs and the corresponding thresholds.

In the CWCS considered in Fig. 1, the matched filter is used to maximize SNR. The filter output $y(t)$ is used to detect the information. In the following, we analyze the return map of $y(t)$ and the corresponding thresholds.

The impulse response of matched filter \cite{Key-18} is $g(t)=p(-t)$ for the chaotic signal generated by Eq. (1). The filter output is
\begin{equation}\label{yt}
\begin{aligned}
y(t)&=g(t)*r(t)\\
&=\int_{\tau=-\infty}^{\infty}p(-\tau)r(t-\tau)d\tau\\
    &=\sum_{l=0}^{L-1}\alpha_l\sum_{m=-\infty}^{\infty}s_m\Big(\int_{\tau=-\infty}^{\infty}p(\tau)p(\tau-t+\tau_l+\frac{m}{f})d\tau\Big)\\
    &+\int_{\tau=-\infty}^{\infty}p(\tau-t)w(\tau)d\tau.
\end{aligned}
\end{equation}
Sampling $y(t)$ at $t=n/f$ we have
\begin{equation}\label{y0}
\begin{aligned}
y_n&=\sum_{l=0}^{L-1}\alpha_l\sum_{m=-\infty}^{\infty}s_m\Big(\int_{\tau=-\infty}^{\infty}p(\tau)p(\tau+\tau_l+\frac{m-n}{f})d\tau\Big)\\
   &+\int_{\tau=-\infty}^{\infty}p(\tau)w(\tau)d\tau=\sum_{l=0}^{L-1}\sum_{m=-\infty}^{\infty}s_mC_{l,m-n}+W\\
    &=\sum_{l=0}^{L-1}s_nC_{l,0}+\sum_{l=0}^{L-1}\sum_{\substack{m \ne {n} \\ m=-\infty}}^{m=\infty}s_mC_{l,m-n}+W\\
    &=s_nP+I+W.
 \end{aligned}
\end{equation}

In Eq. (\ref{y0}), the first term is the expected signal, where $s_n$ is the expected symbol and $P=\sum_{l=0}^{L-1}C_{l,0}$ is the sum of the multipath power for $s_n$. The second term $I$ is the filtered ISI from other symbols. $C_{l,i}=\alpha_l\int_{\tau=-\infty}^{\infty}p(\tau)p(\tau+\tau_l+\frac{i}{f})d\tau$ is calculated using Eq. (\ref{Clm}) on the top of page 5,
\begin{figure*}
\hrulefill 
\begin{equation}\label{Clm}
\begin{aligned}
&C_{l,i}=\\
&\left\{
\begin{aligned}
    &\alpha_lD(2-e^{-\frac{\beta}{f}}-e^{\frac{\beta}{f}})\big(A{\rm{cos}}(\omega\tau_l)+B{\rm{sin}}(\omega\tau_l)\big),\ \ \ \ \     {\rm{if}}\ \ (|\tau_l+\frac{i}{f}|\ge\frac{1}{f})\\
    &\alpha_l\left\{
    \begin{aligned}
    A\big(D(2-e^{-\frac{\beta}{f}})-D^{-1}e^{-\frac{\beta}{f}}\big){\rm{cos}}(\omega\tau_l)+B\big(D(2-e^{-\frac{\beta}{f}})+D^{-1}e^{-\frac{\beta}{f}}\big){\rm{sin}}(\omega\tau_l)+1-|\tau_lf+i|
    \end{aligned}
    \right\},\ \
    {\rm{if}}\ \ (0\le|\tau_l+\frac{i}{f}|<\frac{1}{f})
\end{aligned}
\right.
\end{aligned}
\end{equation}
\hrulefill 
\end{figure*}
where $A=\frac{(\omega^2-3\beta^2)f}{4\beta(\omega^2+\beta^2)}$, $B=\frac{(3\omega^2-\beta^2)f}{4\omega(\omega^2+\beta^2)}$ and $D=e^{-\beta|\tau_l+\frac{i}{f}|}$.

When $\tau_l+i/f\ge1/f$, Eq. (\ref{Clm}) shows that $C_{l,i}$ satisfies
\begin{equation}\label{CliCli1}
\begin{aligned}
C_{l,i+1}/C_{l,i}=e^{-\beta/f}.
\end{aligned}
\end{equation}
By substituting Eq. (\ref{CliCli1}) into Eq. (\ref{y0}), the return map of $y(t)$ can be derived as
\begin{equation}\label{yn1}
\begin{aligned}
y_{n+1}=e^{\frac{\beta}{f}}y_n-\sum_{l=0}^{L-1}\sum_{i=-\infty}^{1}s_{n+i}(e^{\frac{\beta}{f}}C_{l,i}-C_{l,i-1}).
\end{aligned}
\end{equation}
Equation (\ref{yn1}) shows that the return map of filter output is depended on the future symbol $s_{n+1}$, the current symbol $s_n$, and the past symbols $s_{n+i} (i=-\infty,\cdots,-1)$. In fact, it is unnecessary to use all of the past symbols because the coefficient ($e^{\beta/f}C_{l,i}-C_{l,i-1}$) of $s_{n+i}$ is close to 0 for $i<-5-{\tau_l}f$.

Under the two-path channel with $\tau_0=0, \tau_1=1, \gamma=0.9$, the return map of $y(t)$ is given in Fig. 2(d), in which the past symbols $[s_{n-10},s_{n-9},\cdots,s_{n-2}]=[1,-1,1,-1,1,-1,1,-1,1]$ are known. In Fig. 2(d), there are eight branches corresponding to different values of $[s_{n-1}, s_n, s_{n+1}]$, respectively. The branches on the lower half panel represent $s_{n+1}=-1$ and the branches on the top half panel represent $s_{n+1}=1$. The solid lines correspond to $s_n=-1$ and the dashed lines correspond to $s_n=1$, respectively. The red lines correspond to $s_{n-1}=-1$ and the blue lines correspond to $s_{n-1}=1$, respectively. For better distinguishing the different values of $s_n$, the eight branches can be regrouped according to different pairs of $[s_{n-1},s_{n+1}]$. If $[s_{n-1},s_{n+1}]=[-1,-1]$, the branches pair corresponds to the below-red pair, else if $[s_{n-1},s_{n+1}]=[-1,1]$, the branches pair corresponds to the top-red pair, else if $[s_{n-1},s_{n+1}]=[1,-1]$, the branches pair corresponds to the below-blue pair, else (i.e., $[s_{n-1},s_{n+1}]=[1,1]$), the branches pair corresponds to the top-blue pair. Based on Eq. (\ref{yn1}), the optimal JL for $L$ paths channel is
\begin{equation}\label{JLs}
\begin{aligned}
y_{n+1}=e^{\frac{\beta}{f}}y_n-\sum_{l=0}^{L-1}\sum_{\substack{i \ne 0 \\ i=-\infty}}^{1}s_{n+i}(e^{\frac{\beta}{f}}C_{l,i}-C_{l,i-1}).
\end{aligned}
\end{equation}
In the two-path channel and fixed $[s_{n-10},s_{n-9},\cdots,s_{n-2}]=[1,-1,1,-1,1,-1,1,-1,1]$, the JLs for four groups are plotted using dotted lines in Fig. 2(d), and the JD$=2\sum_{l=0}^{L-1}(C_{l,0}-e^{-\beta/f}C_{l,-1})$ in this case. For other combination of past symbols, all the branches and JLs will shift on the plane with the same slope, but the JD is invariant. It is more important to notice that the JD after matched filter is enlarged, i.e., the branches are more separated to get the lower BER as compared to the case using signal $r(t)$ before the matched filter.

For detecting $s_n$, the optimal threshold $\theta_n$ can be set as $y_n$ in the JL, thus
\begin{equation}\label{ThetaN}
\begin{aligned}
\theta_n&=e^{-\frac{\beta}{f}}y_{n+1}+\sum_{l=0}^{L-1}\sum_{\substack{i=-\infty \\ i \ne 0}}^{1}s_{n+i}(C_{l,i}-e^{-\frac{\beta}{f}}C_{l,i-1})\\
&=\sum_{l=0}^{L-1}\sum_{\substack{i \ne {0} \\ i=-\infty}}^{i=\infty}s_{n+i}C_{l,i}=I=I_{past}+I_{future}.
\end{aligned}
\end{equation}

It is quite surprising to see that the right-hand side of Eq. (\ref{ThetaN}) is the filtered ISI, which contains both the past symbols $s_{n+i}(i<0)$ and the future symbols $s_{n+i}(i>0)$. The ISI from past symbols is denoted as
$
    I_{past} = \sum_{l=0}^{L-1}\sum_{i=-\infty}^{-1}s_{n+i}C_{l,i}.
$
If we know the channel parameters, $C_{l,i}$ can be calculated using Eq. (\ref{Clm}). Before detecting $s_n$, the past symbols $s_{n+i}(i=-\infty,\dots,-1)$ have been decoded, thus $I_{past}$ can be calculated. The ISI from future symbols is
$
    I_{future} = K\sum_{i=1}^{\infty}s_{n+i}e^{-\beta{i/f}},
$
where $K=\sum_{l=0}^{L-1}\alpha_l(2-e^{-\beta/f}-e^{\beta/f})e^{-\beta\tau_l}\big(Acos(\omega\tau_l)+Bsin(\omega\tau_l)\big)$ depends on the channel parameters $\alpha_l$ and $\tau_l$. With the assumption that $pr(s_{n+i}=1)=pr(s_{n+i}=-1)=1/2$, where $pr({\cdot})$ is the probability of event `${\cdot}$'. The event $s_{n+i}=1$ means that `$1$' is transmitted, the event $s_{n+i}=-1$ means that `$-1$' is transmitted. $I_{future}$ is an uniform distribution in the range given by $[-|\frac{K}{e^{\beta/f}-1}|,|\frac{K}{e^{\beta/f}-1}|]$. Because $I_{future}$ cannot be calculated at the current time, so $I_{past}$ is a suboptimal threshold based on the available information.

The third term in Eq. (\ref{y0}), $W=\int_{\tau=-\infty}^{\infty}p(\tau)w(\tau)d\tau$, is the filtered noise. If $w(\tau)$ is an AWGN with zero mean, then $W$ is Gaussian noise with zero mean \cite{Key-26}. Assuming that the variance of $W$ is $\sigma_{W}^{2}$, then $y_n$ is a Gaussian random variable with conditional probability distribution
\begin{equation}\label{pdfy0}
p\big(y_n|s_n,I\big)=\frac{1}{\sqrt{{2\pi}\sigma_{W}^{2}}}e^{-{\frac{[y_n-s_nP-I]^2}{2\sigma_{W}^{2}}}}.
\end{equation}

From Eq. (\ref{pdfy0}), the BER using the optimal threshold $\theta_n=I$ for detecting $s_n$ is
\begin{equation}\label{BERL}
    p({\rm{error}}|\theta_n=I)=\frac{1}{2}erfc\Bigg(\frac{P}{\sqrt{2\sigma_{W}^{2}}}\Bigg),
\end{equation}
in which $erfc(\cdot)$ is the complementary error function, $P^2/(2\sigma^{2}_{W})$ is the filtered SNR, where $P$ is defined in Eq. (\ref{y0}). The BER using the suboptimal threshold, i.e., $\theta_n=I_{past}$, for detecting $s_n$ is given as Eq. (\ref{BERIpast}),
\begin{figure*}
\hrulefill 
\begin{equation}\label{BERIpast}
\begin{aligned}
    p({\rm{error}}|\theta_n=I_{past})&=\frac{e^{\beta/f}-1}{4|K|}\int_{-\frac{|K|}{e^{\beta/f}-1}}^{\frac{|K|}{e^{\beta/f}-1}} erfc\Bigg(\frac{P+I_{future}}{\sqrt{2\sigma_{W}^{2}}}\Bigg)d(I_{future})\\
     &=\sqrt{2\sigma_{W}^{2}}\frac{e^{\beta/f}-1}{4|K|}\left\{
    \begin{aligned}
    &z_1\cdot erfc(z_1)-z_2\cdot erfc(z_2)-{e^{-{z_1}^2}}/{\sqrt{\pi}}+{e^{-{z_2}^2}}/{\sqrt{\pi}}\\
    \end{aligned}
    \right\}\\
    \end{aligned}
\end{equation}
\hrulefill 
\end{figure*}
where $z_1=\frac{P+\frac{|K|}{e^{\beta/f}-1}}{\sqrt{2\sigma_{W}^{2}}}$ and $z_2=\frac{P-\frac{|K|}{e^{\beta/f}-1}}{\sqrt{2\sigma_{W}^{2}}}$. Equations (\ref{BERL}) and (\ref{BERIpast}) give the analytical BER using the optimal threshold and the suboptimal threshold, respectively.

\section{Results and discussion}
In the conventional wireless communication system, channel equalization is required to decrease ISI caused by multipath. When both the performance and the algorithm complexity of the equalization are considered, the minimum mean square error (MMSE) equalizer is an excellent choice \cite{Key-28}. In this part, the parameter $\beta=0.65$ and the base frequency $f=1$. To analyze the performance of the proposed CWCS, simulations are performed using different thresholds, including $\theta=0$ (without consideration of multipath effect); $\theta=0$ combined with MMSE; $\theta=I_{past}$ and $\theta=I$ without equalization; The BER lower bound of the single path channel \cite{Key-18} is also plotted in Fig. 3. In the simulation of the proposed method, the retrieved past symbols $s_{n+i} (0\geq i\geq-5+\lceil{-\tau_lf)\rceil}$) were used for calculating the suboptimal threshold $\theta=I_{past}$.
\begin{figure}[H]
\centering{}\includegraphics[scale=0.4]{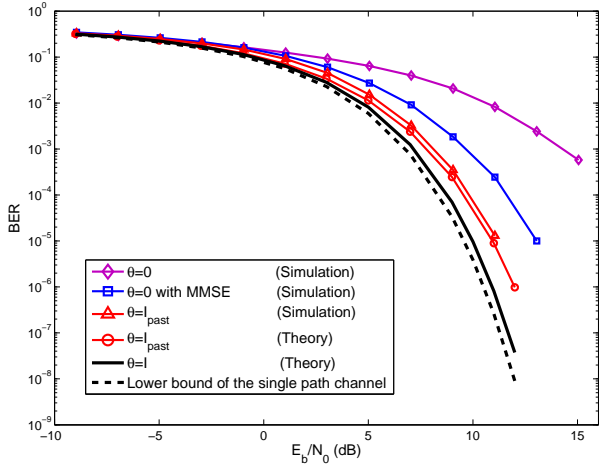}\caption{BER comparison results. Black solid line is the theoretical BER using the optimal threshold $\theta=I$, red solid line with circle markers is the theoretical BER using $\theta=I_{past}$, red solid line with upper triangular markers is the simulation BER using $\theta=I_{past}$, blue solid line with square markers is the simulation result using $\theta=0$ combined with MMSE, violet solid line with diamond markers is the simulation result using $\theta=0$. Black dashed line is the theoretical BER lower bound for the single path channel.}
\end{figure}
\begin{figure}[tbh]
\centering{}\includegraphics[scale=0.4]{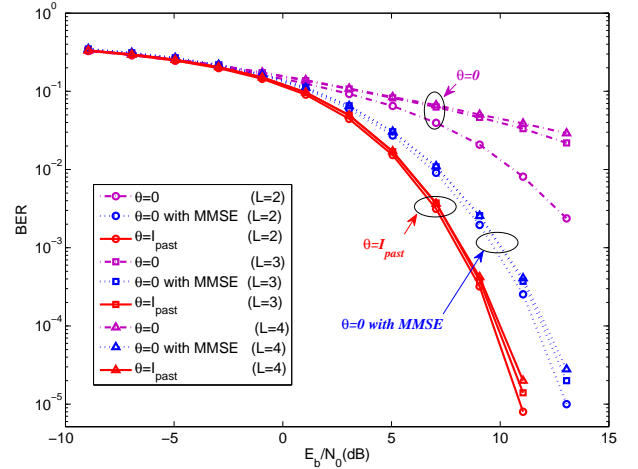}\caption{BER comparison results under different number of paths. The red solid lines are the BER curves using the suboptimal threshold $\theta=I_{past}$, the blue dotted lines are the BER curves using $\theta=0$ combined with MMSE, and the violet dash-dotted lines are the BER curves using $\theta=0$. The lines with circle markers are the BER curves for $L=2$, the lines with square markers are the BER curves for $L=3$, and the lines with upper triangular markers are the BER for $L=4$.}
\end{figure}

In Fig. 3, we assume that the energy cost perbit is $E_b=1$, two paths with propagation delay $\tau_0=0,\tau_1=1$ and damping coefficient $\gamma=0.6$. The simulation results are obtained by averaging over 50,000 trials. From Fig. 3, we learn that using $\theta=0$, the CWCS has the worst BER, because the multipath effect is completely ignored under such condition. BER is reduced by introducing MMSE equalization, but using suboptimal threshold, i.e., $\theta=I_{past}$ without MMSE, BER is not only lower than those obtained by the above two methods, but also close to the optimal threshold $\theta=I$ and to the theoretical lower bound of single path channel with respect to different $E_b/N_0$. Therefore, the CWCS using the suboptimal threshold decreases the BER efficiently as compared to the method using $\theta=0$ with conventional MMSE equalizer.

\begin{table*}
 \centering
   \caption{Required $E_b/N_0$ (dB) for BER = $10^{-3}$ over inaccurate channel parameters}
   \begin{tabular}{|c|c|p{1.5cm}<{\centering}|p{1.5cm}<{\centering}|p{1.5cm}<{\centering}|p{1.5cm}<{\centering}|}
    \hline
     Path number &Required $E_b/N_0$ (dB) &$\varepsilon=0$ &$\varepsilon=0.1$ &$\varepsilon=0.2$ &$\varepsilon=0.3$\\
     \cline{1-6}
     \multirow{2}{*}{L=2} & $\theta=I_{past}$ (simu.) &8.04 &8.09 &8.26 &8.51\\
     \cline{2-6}
     &$\theta=0$ with MMSE (simu.) & 9.59 & 9.74 & 10.16 & 10.33\\
     \hline
      \multirow{2}{*}{L=3} & $\theta=I_{past}$ (simu.) &8.15 &8.22 &8.51 &8.77\\
     \cline{2-6}
     &$\theta=0$ with MMSE (simu.)& 9.98 & 10.23 & 11.28 & 12.87\\
     \hline
   \end{tabular}
 \end{table*}

Considering the wireless channel in practice with more multipaths, Fig. 4 gives the BER comparison results under different number of multipaths with channel fading $h(\tau_l)=e^{-0.6\tau_l}$, where $L$ is the number of multipaths. From Fig. 4, we can see that, for the wireless channel with more number of multipaths, the performance of the proposed suboptimal threshold $\theta=I_{past}$ is better as compared with that of $\theta=0$ with MMSE equalization.

The BER comparison results between the proposed method and the conventional system with binary phase-shift keying (BPSK) modulation are shown in Fig. 5. In the conventional system, MMSE channel equalization was used for mitigating the ISI caused by multipath propagation. From Fig. 5, we find that the BER of the proposed method is very close to that of the traditional BPSK for the single path channel. However, in the multipath channel, the proposed method is better than the traditional BPSK with MMSE equalization.
\begin{figure}[tbh]
\centering{}\includegraphics[scale=0.4]{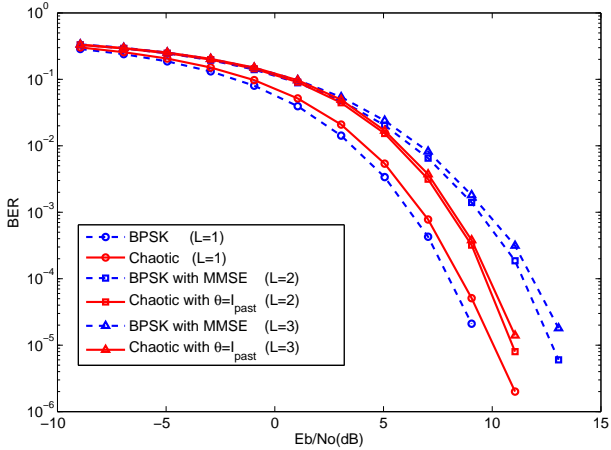}\caption{BER comparison results between the proposed method and the conventional BPSK with MMSE. The red solid lines are the BER curves of the proposed method using the suboptimal threshold $\theta=I_{past}$, and the blue dashed lines are the BER curves of traditional communication system using BPSK modulation and MMSE channel equalization. The lines with circle markers are the BER curves for the single path channel ($L=1$), the lines with square markers are the BER curves for the two-path channel ($L=2$) with $\tau_0=0,\tau_1=1,\gamma=0.6$, and the lines with upper triangular markers are the BER curves for the three-path channel ($L=3$) with $\tau_0=0,\tau_1=1,\tau_2=2,\gamma=0.6$.}
\end{figure}

In the CWCS, the accurate channel parameters are required for equalization and for calculating the threshold $\theta$. In the following, the effects of inaccurate channel parameters, due to the simplified channel model, are analyzed. We assume that the used channel $\hat{h}=h+\Delta h$ is different from the accurate channel $h$, where $\Delta h$ is the channel error uniformly distributed in the range $[-\varepsilon h, \varepsilon h]$. For two-path channel with $\tau_0=0,\tau_1=1,\gamma=0.6$ and three-path channel with $\tau_0=0,\tau_1=1,\tau_2=2,\gamma=0.6$, Table 1 gives the required $E_b/N_0$ for BER $=10^{-3}$ under different $\varepsilon$. In Table 1, two methods, $\theta=I_{past}$ and $\theta=0$ with MMSE, are given for comparison. The simulation results show that the BER performances of both methods are affected by the channel error, the higher $\varepsilon $ and the higher required $E_b/N_0$ due to the inaccurate channel parameter $\hat{h}$ are used. For the method of $\theta=0$ with MMSE, the required $E_b/N_0$ is increased by 0.74dB/2.89dB for the two/three-path channel with error $\varepsilon=0.3$. However, for the proposed suboptimal threshold, i.e., $\theta=I_{past}$, the required $E_b/N_0$ is only increased by 0.47dB/0.62dB for two/three-path channel with the same channel error. The result shows that the proposed suboptimal threshold is robust to channel error, and can efficiently decrease BER caused by the multipath propagation. As compared to $\theta=0$ with MMSE, the proposed method is better.
\begin{figure}[tbh]
\centering{}\includegraphics[scale=0.39]{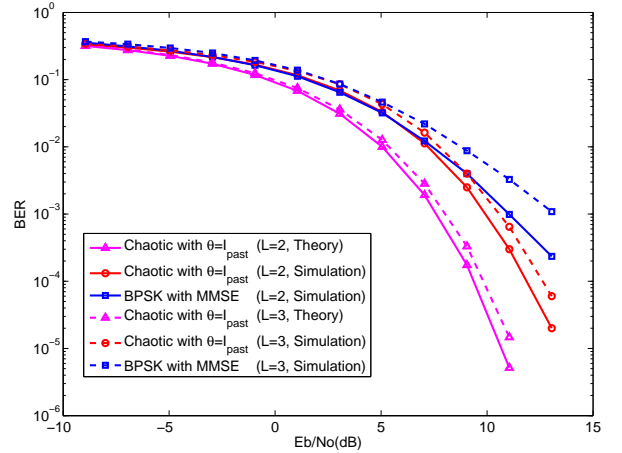}\caption{{\color{blue}BER comparison results between the proposed method and the conventional BPSK with MMSE in a time-varying wireless channel. In this simulation, the channel damping coefficient $\gamma$ is randomly changed for different frames. The channel parameters estimation by the LS algorithm are used to implement the comparison of both methods. The pink lines are the theoretical BER curves of the proposed method using the suboptimal threshold $\theta=I_{past}$, and the red lines are the simulation BER curves of the proposed method using $\theta=I_{past}$, and the blue lines are the simulation BER curves of the traditional communication system using BPSK modulation and MMSE channel equalization. The solid lines are the corresponding BER curves for the two-path channel ($L=2$), and the dashed lines are the corresponding BER curves for the three-path channel ($L=3$).}}
\end{figure}

{\color{blue}
In the above, the channel is static during the information transmission. However, in practice, the wireless channel is time-varying due to frequency-selection, Doppler effect and other factors. In such a case, the wireless channel can be described as a time-varying but quasi-static channel, where the channel parameters are assumed to be unchanged within one frame, but are varied from one frame to the next. Figure 6 gives the BER comparison results under a quasi-static channel. In the simulation, the channel damping coefficient $\gamma$ is a uniformly distributed random variable in the interval [0.3, 0.9] for the different frames. There are 4096 bits in one frame, which contains 256 training bits and 3840 data bits. The training bits are used for channel parameters estimation by the least squares (LS) algorithm. The estimated channel parameters are used to implement our method and MMSE algorithm for the conventional BPSK system. The simulation results are obtained by averaging over 2000 frames. In Fig. 6, BER curves of the proposed CWCS, conventional BPSK system, and the theoretical results in Eq. (\ref{BERIpast}) are given for comparison. We can see that the BERs of both methods are worse than the results in Fig. 5, because of the inaccurate channel estimation. However, for both two and three-path channels, the proposed CWCS is better than the conventional BPSK with MMSE equalization.
}

It is worth noting that, in the proposed method, the past symbols are needed to regroup the return map. In the simulation, the first $5-\lceil{-\tau_lf\rceil}$ symbols of the transmitted sequences are known for initial regrouping. In the practical communication system, the known preamble signal is often used to estimate the channel parameters and to acknowledge the receiver that the communication will start soon, the preamble signal can be used to implement the initial regrouping of the proposed method.

\section{Conclusions}
To summarize, the influence of multipath on information decoding using the return map of chaotic signals is reported in this paper. We find that the multi-branch return map of the received signal (and the filtered signal) can be regrouped and the different $judgment\ lines$ (JL) can be used to decrease BER caused by the multipath propagation. We also find that, after the matched filter, the JD is increased to decrease BER. We obtain the optimal threshold $\theta=I$, which is surprisingly the ISI for the filtered signal. Using the optimal threshold, BER is very close to the lower bound of BER for the single path channel. Unfortunately, the optimal threshold needs the future information symbols, which are unavailable at the current time. To deal with such a situation, we proposed a suboptimal threshold using the past symbols. The analytical expression of BER using suboptimal threshold is derived, which contains the channel parameters and the past symbols, and thus can be obtained by channel estimation and symbols detected in the past. Simulation and numerical analysis results show that the suboptimal threshold can eliminate most of multipath effects, and it works well even in the practical wireless channel with uncertain parameters. The results in this paper, together with the results in \cite{Key-22}, show the merits of CWCS, including: i) a simple encoding method consumes less energy using impulse control \cite{Key-23,Key-24,Key-25}; ii) noise-like signal is transmitted (it is difficult for the intruder without the system knowledge to retrieve the information); iii) a simple decoding threshold with less computation cost can resist the multipath effects and achieve a highly competitive BER performance.
\section*{Acknowledgments}
This work is supported by NSFC under Grant No. 61401354, No. 61172070, and No. 61502385; by the Innovative Research Team of Shaanxi Province under Grant No. 2013KCT-04; and by Key Basic Research Fund of Shaanxi Province under Grant No. 2016JQ6015.

\bibliographystyle{apsrev4-1}
\bibliography{mybibfile}

\end{document}